\documentstyle[thmsa,11pt,a4,sw20lart]{article}

\input tcilatex
\QQQ{Language}{
American English
}

\begin{document}

\topmargin 0pt \oddsidemargin 5mm

\setcounter{page}{1}

\hspace{8cm}{} \vspace{2cm}

\begin{center}
{\large {CORRELATED FAST ION STOPPING IN MAGNETIZED CLASSICAL PLASMA}}\\

H.B. Nersisyan$^{(1)}$ and C. Deutsch$^{(2)}$\\\vspace{1cm} $^{(1)}$ {\em %
Division of Theoretical Physics, Institute of Radiophysics and Electronics,
Ashtarak-2, 378410, Republic of Armenia}\\

$^{(2)}$ {\em Laboratoire de Physique des Gaz et Plasmas, Universit\'e Paris
XI, B\^at.10, 91405 Orsay, France}
\end{center}

\vspace {5mm} \centerline{{\bf{Abstract}}}

The results of a theoretical investigation on the stopping power of ion pair
in a magnetized electron plasma are presented, with particular emphasis on
the two-ion correlation effects. The analysis is based on the assumptions
that the magnetic field is classically strong ($\lambda _B\ll a_c\ll \lambda
_D$, where $\lambda _B$, $a_c$ and $\lambda _D$ are respectively the
electron de Broglie wavelength, Larmor radius and Debye length) and that the
velocity of the two ions is identical and fixed. The stopping power and {\it %
vicinage function} in a plasma are computed by retaining two-ion correlation
effects and is compared with the results of the individual-projectile
approximation.

\vspace {5mm} PACS number(s): 52.40.Mj, 52.35.-g \newpage 

The problem of the interaction of a beam of fast charged particles with a
plasma has been attracting more and more interest in the last years in
connection with the proposed scheme of inertial confinement fusion based on
the use of heavy-ion beams to drive the $DT$ target towards the ignition
conditions [1].

Several physical situations can be conceived in which the beam ions are
closely spaced so that their stopping in the ionized medium is not exactly
that of charged particles whose dynamics is independent of the presence of
each other, but suffers from mutual correlations [2, 3]. It can be the case
of very high-density ion beams or, more realistically, when ion clusters are
to be used instead of standard ion beams [4].

The nature of experimental plasma physics is such that experiments are
usually performed in the presence of magnetic fields, and consequently it is
of interest to investigate the effects of a magnetic field on the stopping
power. The strong magnetic fields used in laboratory investigations of
plasmas can appreciably influence the processes determined by Coulomb
collisions [5]. This influence is even more important in white dwarfs and in
neutron stars, the magnetic fields on the surfaces of which can attain
strengths up to $10^5$-$10^{10}kG$.

The stopping of uncorrelated charged particles in a magnetized plasma has
been the subject of several papers [6-9]. The stopping of a fast test
particle moving with a velocity $u$ much higher than the electron thermal
velocity $v_T$ was studied in [6, 8]. The energy loss of a charged particle
moving with arbitrary velocity was studied in [7]. The expression derived
there for the Coulomb logarithm corresponds to the classical description of
collisions.

In ref. [9] expressions were derived describing the stopping power of a slow
charged particle in Maxwellian plasma with a classically strong (but not
quantizing) magnetic field ($\lambda _B\ll a_c\ll \lambda _D$, where $%
\lambda _B$, $a_c$ and $\lambda _D$ are respectively the electron de Broglie
wavelength, Larmor radius and Debye length), under conditions such that the
scattering processes must be described quantum-mechanically.

Here we present an evaluation of the stopping power of a ion pair moving in
a warm collisionless plasma placed in a classical strong homogeneous
magnetic field, with the aim to show under which conditions correlation
effects can be important and how they modify the single-particle expression
of the stopping power.

A uniform plasma is considered in the presence of a homogeneous magnetic
field $B_0$ which is assumed sufficiently strong so that $\lambda _B\ll
a_c\ll \lambda _D$. From these conditions we can obtain $3\cdot
10^{-6}n_0^{1/2}<B_0<10^5T$ ($n_0$ and $T$ are respectively unperturbed
number density and temperature of plasma), where $n_0$ is measured in $%
cm^{-3}$, $T$ is measured in $eV$ and $B_0$ in $kG$. Because of this
assumption, the perpendicular cyclotron motion of the ions and plasma
electrons is neglected.

Let us analyze the interaction process of a two-ion system moving in plasma.
The interionic vector is ${\bf l}$. Also, due to the high frequencies
involved, the very weak response of the plasma ions is neglected and the
Vlasov-Poisson equations to be solved for the perturbation to the electron
distribution function, $f_1$, and the potential $\varphi $, are as follows:

\begin{equation}
\left( \frac \partial {\partial t}+({\bf bv})({\bf b}\nabla )\right) f_1(%
{\bf r},{\bf v},t)=-\frac em({\bf b}\nabla \varphi )\left( {\bf b}\frac{%
\partial f_0(v)}{\partial {\bf v}}\right) ,
\end{equation}
\begin{equation}
\nabla ^2\varphi =-4\pi e\left[ Z_1\delta ({\bf r}-{\bf l}-{\bf u}%
t)+Z_2\delta ({\bf r}-{\bf u}t)\right] +4\pi e\int d{\bf v}f_1({\bf r},{\bf v%
},t),
\end{equation}
where $b$ is the unit vector parallel to $B_0$, $f_0$ is the unperturbed
electron distribution function which is taken to be uniform and Maxwellian, $%
{\bf u}$ is the velocity of two ions. Here, $Z_1e$ and $Z_2e$ are the
effective charges of the two ions, assumed to be constant throughout the
slowing-down process.

By solving the eqs. (1) and (2) in space-time Fourier components, we obtain
the following expression for the electrostatic potential:

\begin{equation}
\varphi ({\bf r},t)=\frac{4\pi e}{(2\pi )^3}\int d{\bf k}\frac{\exp [i{\bf k}%
({\bf r}-{\bf u}t)]}{k^2\varepsilon ({\bf k},{\bf ku})}\left[ Z_2+Z_1\exp (-i%
{\bf kl})\right] ,
\end{equation}
where

\begin{equation}
\varepsilon ({\bf k},\omega )=1+\frac 1{k^2\lambda _D^2}W\left( \frac \omega
{|{\bf kb}|v_T}\right) 
\end{equation}
is the longitudinal dielectric function for a Maxwellian plasma placed in a
strong magnetic field and $W(\xi )=A(\xi )+iB(\xi )$ is the plasma
dispersion function [10].

The stopping power of the ion pair can be computed by summing up the forces
acting on two ions, due to the electric field induced in the plasma; it reads

\begin{equation}
S=-\frac{dW}{dz}=\left( Z_1^2+Z_2^2\right) S_{{\rm ind}}+2Z_1Z_2S_{{\rm corr}%
}({\bf l}),
\end{equation}
where

\begin{equation}
S_{{\rm ind}}=\frac{4\pi e^2}{(2\pi )^3u}\int d{\bf k}\frac{{\bf ku}}{k^2}%
{\rm Im}\frac{-1}{\varepsilon ({\bf k},{\bf ku})},
\end{equation}
\begin{equation}
S_{{\rm corr}}=\frac{4\pi e^2}{(2\pi )^3u}\int d{\bf k}\frac{{\bf ku}}{k^2}%
{\rm Im}\frac{-1}{\varepsilon ({\bf k},{\bf ku})}\cos ({\bf kl}).
\end{equation}

There are two contributions to the stopping power of two ions. The first one
is the uncorrelated particle contribution and represents the energy loss of
the two projectiles due to the coupling with collective plasma modes (the
first term in eq. (5)). The second contribution is responsible for the
correlated motion of two ions by means of the resonant interaction with the
excited plasma waves (the second term in eq. (5)). Both terms are
responsible of the irreversible transfer of the two-ion energy to plasma
through resonant electrons.

The stopping properties of the medium can be also conveniently described in
terms of the {\it partial range} defined as follows:

\begin{equation}
R(E)=\frac{E_{{\rm in}}-E}{S(E_{{\rm in}})},
\end{equation}
where $E_{{\rm in}}$ is the injection ion energy and $E$ is the actual value
of the ion energy during the slowing-down process. For $E\cong 0$, $R(0)$
gives the penetration depth ({\it range}) of the incident particle.

Let us begin with the evaluation of the stopping power of two ions with the
same effective charge $Ze$. For an arbitrary relative position of the two
test ions, the expressions of the uncorrelated (proportional to $S_{{\rm ind}%
}$) and correlated (proportional to $S_{{\rm corr}}$) stopping powers of the
ion pair becomes:

\begin{eqnarray}
S_{{\rm ind}}(\lambda ) &=&\frac{e^2}{2\pi \lambda _D^2}\left\{ \frac{%
B(\lambda )}2\ln \frac{B^2(\lambda )+\left( s^2+A(\lambda )\right) ^2}{%
B^2(\lambda )+A^2(\lambda )}-\right.  \\
&&\left. -A(\lambda )\left[ \arctan \frac{s^2+A(\lambda )}{B(\lambda )}%
-\arctan \frac{A(\lambda )}{B(\lambda )}\right] \right\} ,
\end{eqnarray}
\begin{equation}
S_{{\rm corr}}(l,\vartheta ,\lambda )=\frac{2e^2}{\pi \lambda _D^2}B(\lambda
)\int_0^s\frac{k^3dk}{[k^2+A(\lambda )]^2+B^2(\lambda )}Q(kL\cos \vartheta
;kL\sin \vartheta ).
\end{equation}
Here

\begin{equation}
Q(a;b)=\int_0^1\cos (ax)J_0\left( b\sqrt{1-x^2}\right) xdx,
\end{equation}
$J_0(z)$ is the Bessel function of the zero order, $\vartheta $ is the angle
between the interionic vector ${\bf l}$ and the velocity vector ${\bf u}$, $%
L=l/\lambda _D$, $\lambda =u/v_T$, $s=k_{\max }\lambda _D$ with $k_{\max
}=1/r_{\min }$, where $r_{\min }$ is the effective minimum impact parameter.
Here $k_{\max }$ has been introduced to avoid the divergence of the integral
caused by the incorrect treatment of the short-range interactions between
the ion pair and the plasma electrons within the linearized Vlasov theory.
The value of $k_{\max }$, will be $1/a_c$ for fusion plasmas, since the
magnetized plasma approximation which neglects the perpendicular motion of
the electrons ceases to be valid for collision parameters less than $a_c$.

Consistently with the notation introduced above we separate the
single-particle contribution from the correlated one of the stopping power.
Then, defining the {\it interference} or {\it vicinage function} $\chi ({\bf %
l})$ as [2, 3]

\begin{equation}
\chi (l,\vartheta ,\lambda )=\frac{S_{{\rm corr}}(l,\vartheta ,\lambda )}{S_{%
{\rm ind}}(\lambda )},
\end{equation}
expression (5) can be put in the form

\begin{equation}
S=2Z^2S_{{\rm ind}}(\lambda )\left( 1+\chi ({\bf l})\right) .
\end{equation}
Here, $\chi $ describes the intensity of correlation effects with respect to
uncorrelated situation. In the case of the fast ions ($\lambda \gg 1$), from
expressions (9) and (10) we obtain:

\begin{equation}
\chi (l,\vartheta ,\lambda )\cong 2Q\left( \frac L\lambda \cos \vartheta
;\frac L\lambda \sin \vartheta \right) .
\end{equation}
For two values of the orientation angle $\vartheta =0^0$ and $\vartheta
=90^0 $ the expression (14) for {\it vicinage} function becomes [11]:

\begin{equation}
\chi (l,0,\lambda )=2\left( \frac{\sin (L/\lambda )}{(L/\lambda )}-\frac{%
1-\cos (L/\lambda )}{(L/\lambda )^2}\right) ,
\end{equation}
\begin{equation}
\chi (l,\pi /2,\lambda )=\frac 2{(L/\lambda )}J_1(L/\lambda ),
\end{equation}
where $J_1(z)$ is the Bessel function of the first order.

Fig. 1 shows $\chi $ as a function of $L$ and $\vartheta $ for $\lambda =10$
and $s=10$. It is shown that $\chi $ can decrease or increase for large $%
\vartheta $-value, depending on the interionic distance $l$ (see also
expressions (15) and (16)). Meanwhile, the correlation effects decrease for
large $\vartheta $-value in plasma in the absence of magnetic field [2]. It
should be clear that this effect is accounted by the character of
electrostatic potential of test charged particles in magnetized plasma. As
shown in [12], in the frame of the test particle, moving in a plasma placed
in the strong magnetic field ($a_c<\lambda _D$), the part of spatially
oscillatory potential has spherical symmetry over the hemisphere behind the
particle and is zero ahead of the particle. The second part has different
character, which makes the potential continuous at the plane containing the
particle, is oscillatory in the radial direction, but decreases almost
monotonically in the axial direction.

Figs. 2 and 3 show $\chi $ as a function of $\lambda $, in the cases of $%
\vartheta =0^0$ and $\vartheta =90^0$ respectively, for different values of $%
L$. As expected, correlation effects increase in the high-velocity limit or
for small $L$-values. In the cases of $\vartheta =15^0$, $\vartheta =45^0$
and $\vartheta =75^0$, the correlation effect between two protons is shown
in Fig. 4, where the total range $R(0)$ is shown as a function of $L$. The
parameters are: $E_{{\rm in}}=9.2MeV$, $T=100eV$, $n_0=10^{22}cm^{-3}$, $%
B_0=10^6kG$ (such conditions are possible on the surface of a neutron star).
The halving of $R(0)$ (corresponding to the doubling of the stopping power),
with respect to the case of two uncorrelated protons, is shown. It should be
clear that test-ion densities corresponding to $l\cong u/\omega _p$ are
unrealistic for conventional ion beams. However, small interionic distances (%
$\cong 10^{-8}cm$) occur in clusters [4].

\newpage\ 

\begin{center}
{\bf Figure Captions}
\end{center}

Fig.1. The {\it vicinage} function $\chi $ {\it vs}. $L=l/\lambda _D$ and $%
\vartheta $ (degrees). The parameters are: $\lambda =10$, $s=10$.

Fig.2. The {\it vicinage} function $\chi $ {\it vs}. $\lambda $ for $%
\vartheta =0^0$ and $s=10$. Dashed line: $L=1$; dotted line: $L=11$; solid
line $L=21$.

Fig.3. The {\it vicinage} function $\chi $ {\it vs}. $\lambda $ for $%
\vartheta =90^0$ and $s=10$. Dashed line: $L=1$; dotted line: $L=11$; solid
line $L=21$.

Fig.4. Total range $R$ (cm) {\it vs}. $L$ for two protons ($E_{{\rm in}%
}=9.2MeV$) moving in a magnetized electron plasma ($T=100eV$, $%
n_0=10^{22}cm^{-3}$, $B_0=10^6kG$). Solid line: $\vartheta =15^0$; dashed
line: $\vartheta =45^0$; dotted line: $\vartheta =75^0$; horizontal solid
line: uncorrelated protons.

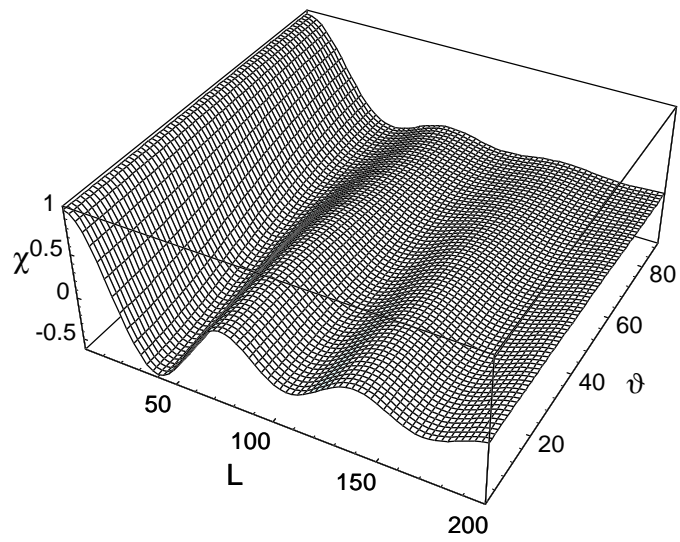
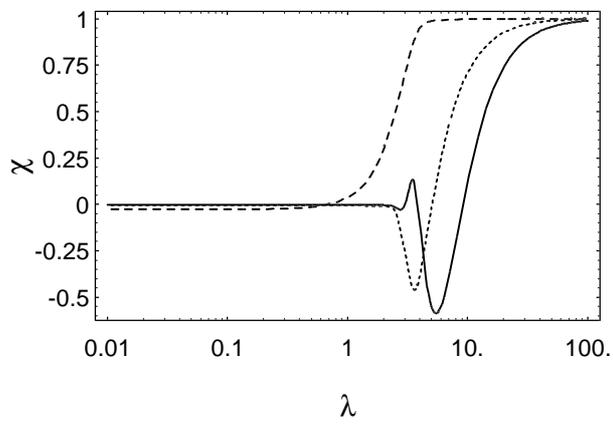
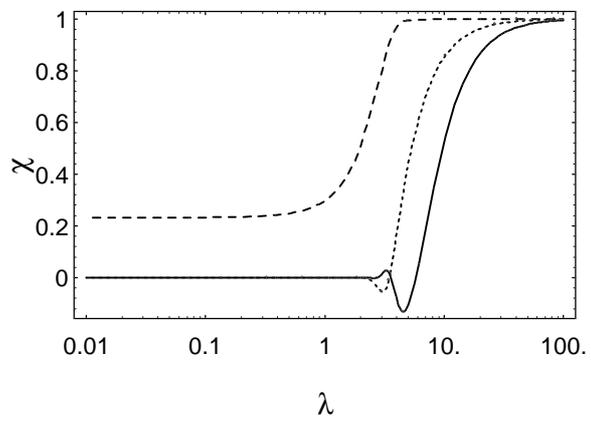
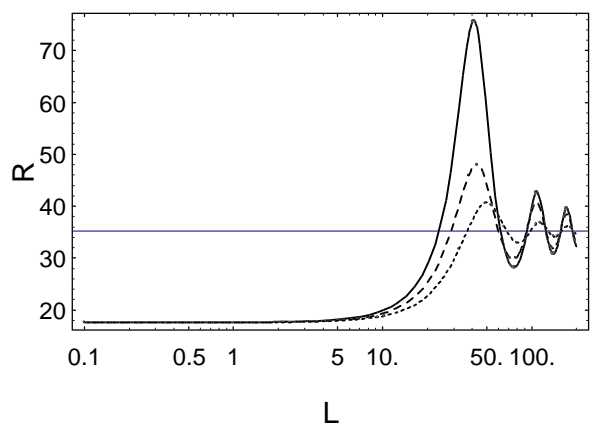
\end{document}